\documentclass{elsart}
\usepackage{graphicx,amssymb}
\journal{Computer Physics Communications}
\usepackage{alltt}
\usepackage{multirow}

\newcommand{\ADF}{\textsc{ADF95}}
\newcommand{\FORTRAN}{\textsc{FORTRAN}}
\newcommand{\FORTRANNF}{\textsc{FORTRAN~90/95}}
\newcommand{\FORTRANN}{\textsc{FORTRAN~90}}
\newcommand{\FORTRANF}{\textsc{FORTRAN~95}}
\newcommand{\FORTRANS}{\textsc{FORTRAN~77}}
\newcommand{\FORTRANSF}{\textsc{FORTRAN~77/90/95}}
\newcommand{\FORTRAND}{\textsc{FORTRAN~2003}}
\newcommand{\FN }{\textsc{F90}}
\newcommand{\FFN}{\textsc{F95}}
\newcommand{\AUTODERIV}{\textsc{AUTO\_DERIV}}
\newcommand{\AUTODERIVS}{\textsc{\tiny AUTO\_DERIV}}
\newcommand{\keyw }[1]{\texttt{#1}}
\newcommand{\keywb}[1]{{\bf #1}}
\newcommand{\startcode}{\begin{alltt}\begin{tabbing}}
\newcommand{\stopcode }{\end{tabbing}\end{alltt}}
\begin{document}
\begin{frontmatter}
\title{ADF95: Tool for automatic differentiation of a \textsc{FORTRAN} code
designed for large numbers of independent variables}
\author{Christian W. Straka} 
\address{Institut f\"ur Theoretische Astrophysik, Universit\"at Heidelberg, Tiergartenstra{\ss}e 15, 69121 Heidelberg, Germany}
\ead{cstraka@ita.uni-heidelberg.de}


\begin{abstract}
\ADF{} is a tool to automatically calculate numerical first derivatives
for any mathematical expression as a function of
user defined independent variables. Accuracy of
derivatives is achieved within machine precision.
\ADF{} may be applied to
any \FORTRANSF{} conforming code and requires minimal changes by
the user. It provides a new derived data type that holds the
value and derivatives and applies forward differencing by
overloading all \FORTRAN{} operators and intrinsic functions. An
efficient indexing technique leads to a reduced
memory usage and a substantially increased performance gain over
other available tools with operator overloading. This gain
is especially pronounced for sparse systems with large number of
independent variables. A wide class of numerical simulations, 
e.g., those employing implicit solvers, can profit from \ADF{}.
\end{abstract}

\end{frontmatter}

{\bf PROGRAM SUMMARY}

{\bf Nature of problem:}

In many areas in the computational sciences first order partial
derivatives for large and complex set of equations are needed
with machine precision accuracy. For example, any implicit or
semi-implicit solver requires the computation of the Jacobian matrix,
which contains the first derivatives with respect to the
independent variables. \ADF{} is a software module to facilitate
the automatic computation of the first partial  derivatives
of any arbitrarily complex mathematical \FORTRAN{} expression. The
program exploits the sparsity inherited by many set of equations
thereby enabling faster computations compared to alternate [1]
differentiation tools.

{\bf Solution method:}

A class is constructed which applies the
chain rule of differentiation to any \FORTRAN{} expression, to compute
the first derivatives by forward differencing. An
efficient indexing technique leads to a reduced
memory usage and a substantially increased performance gain when
sparsity can be exploited. From a users point of view, only
minimal changes to his/her original code are needed in order
to compute the first derivatives of any expression in the code.

{\bf Restrictions:}

Processor and memory hardware may restrict both the
possible number of independent variables and the computation time.

{\bf Unusual features:}

\ADF{} can operate on user code that makes use of the
array features introduced in \FORTRANN{}.
A convenient extraction subroutine for the Jacobian matrix is
also provided.

{\bf Running time:}

In many realistic cases, the
evaluation of the first order derivatives of a
mathematical expression is only six times slower compared
to the evaluation of analytically derived and hard-coded
expressions. The actual factor depends on the
underlying set of equations for which derivatives are to be
calculated, the number of independent variables, the
sparsity and on the FORTRAN 95 compiler.

{\bf References:}

[1] S.Stamatiadis, R.Prosmiti, S.C.Farantos, Comp. Phys. Commun. 127 (2000) 343.

\clearpage
{\bf LONG WRITE-UP}

\section{Introduction}
\ADF{} is a software module to facilitate the analytic
computation of the {\em first\/} partial  derivative of any arbitrarily
complex mathematical \FORTRAN{} expression
including user defined and/or intrinsic functions and subroutines. The
derivatives are computed with respect to independent variables
which must be specified by the user. It must be emphasised that
\ADF{} does not provide the analytic derivative in functional form, rather
it computes the numerical values of the analytic derivatives.
\ADF{} references its computed and internally stored derivatives
with an indexing technique which results in reduced memory usage
of sparse systems.
Thereby it enables faster computations in many practical applications with
large numbers of independent variables.

In many areas in the computational sciences the phenomena to be
simulated can be approximated by solving systems of coupled differential
equations. A quite general class of differential equations, e.g., is the
following initial value problem:
\begin{equation}
B[\vec{y}(t),t]\cdot \vec{y}'(t) = \vec{f}[\vec{y}(t),t]\mbox{,}\quad\quad \vec{y}(t_{0}) = \vec{y}_{0}\mbox{,}
\label{eq:ivp}
\end{equation}
where $\vec{y}(t)$ denotes a $n$-dimensional vector, $\vec{f}$ an arbitrary
$n$-dimensional vector valued
function and $B$ a $n \times n$ matrix. $\vec{y}(t)$ is called a
{\em solution\/} in the interval $I=[t_{0},t_{E}]$
if Eq.(\ref{eq:ivp}) is fulfilled for all $t \in I$. Any
{\em implicit\/} solution strategy requires the computation of
the $n \times n$ Jacobian matrix of the residual:
\begin{equation}
J = \frac{\partial}{\partial \vec{y}} (\vec{f} - B \cdot \vec{y}'(t))\mbox{.}
\end{equation}
Thus the Jacobian contains the first derivatives of the
residual with respect to the independent vector variable $\vec{y}$.
The need for an convenient albeit accurate determination of
first derivatives for the class of implicit solvers has driven my
motivation to develop \ADF. However,
\ADF{} may be useful in all instances where an automatic, efficient
and to working precision
accurate generation of first derivatives are needed.
Only minimal changes in user code are required. 

The functionality of \ADF{} can only be achieved by making use of the new
\FORTRANF{} (\FFN) features \cite{fortran95}
that allow for object-oriented programming. By
defining a new compound variable of {\em derived type}, and re-defining
operators and functions that act on these types with the mechanism of
{\em operator overloading\/} within the encapsulation mechanism
provided by {\em modules\/} we construct a class which applies the
chain rule of differentiation to any \FORTRAN{} expression to compute
the first derivatives by
{\em forward differencing}. All overloaded operators and functions
are defined as {\em elemental\/} and
can thus be called with array arguments of any rank. This is extremely
useful for codes that make use of the array capabilities introduced
in \FN{} \cite{fortran90} and may help compilers to vectorise
or parallelise the code.

A growing number of tools exist \cite{autodiff} for the task of
automatically computing derivatives of \FORTRAN{} expressions.
Among them, two different approaches can be distinguished. The first
method operates on the source code itself generating new source code
for the derivatives. Both initial and generated code must be
compiled in a subsequent invocation of the compiler. The advantage
of this approach lies in the production of generally faster
executables for the differentiation task. It is possible to
use both the forward and the reverse mode of automatic differentiation.
Disadvantages of the latter are that
new code must be generated for any slight change in the parent code.
Furthermore it is more difficult to pass the calculated derivatives
of subroutines to the calling routine. Moreover, new language
features are more difficult to add to these tools.

The second method
makes use of operator overloading. The disadvantages of this
method are the advantages of the source code approach and vice versa.
NAG \cite{nagad} is working on a solution that attempts to combine the
advantages of source code transformation with operator overloading
by adding new compiler functionality. While potentially
exciting, code portability is lost. \ADF{} is conceptually similar
to \AUTODERIV{} {\cite{stamatiadis00}} which, in addition, can
provide second derivatives. In contrast to \AUTODERIV{}
no modification of code utilising array notation is needed
with \ADF. The main enhancement of \ADF{} over existing
approaches with operator overloading is its internal, indexed
storage method that allows more efficient execution in case of
sparse systems with large numbers of independent variables.

\section{\textsc{FORTRAN~90/95} concepts}
A brief summary of concepts introduced in \FORTRAN{} by the two
major revisions \cite{fortran95,fortran90} and used in \ADF{}
is given in this section. A thorough explanation of \FORTRAN{}
language usage can be found, e.g., in the books~\cite{adams97,metcalf00}.

The current standard allows to define new data types in addition to
the built-in ones (\keyw{integer}, \keyw{real}, etc.). These
{\em derived types\/} constitute aggregates of built-in and/or other
derived types. For example, the following derived type
\startcode
\quad \=\quad \kill
\keywb{type} vector     \\
\> \keywb{real} :: x, y  \\
\keywb{end type} vector
\stopcode
defines a new data structure that may represent
a 2-dimensional vector. Whereas the compiler ``knows" how to
perform a mathematical operation on built-in types, it cannot
possibly know how to apply those to derived types. The programmer
can give a meaning to an operation between derived types by, first,
defining a new function, and secondly, {\em overloading\/}
the operator with this function. The following code provides
the functionality for adding two variables of \keyw{type(vector)}
employing the rules of vector calculus
\startcode
\quad \=\phantom{\keywb{type}(vector), \keywb{intent}(in)} \=\quad \kill
\keywb{function} vadd(v, w) \\
\> \keywb{type}(vector), \keywb{intent}(in) \> :: v, w \\
\> \keywb{type}(vector) \> :: vadd \\

\> vadd\%x = v\%x + w\%x \\
\> vadd\%y = v\%y + w\%y \\
\keywb{end function} vadd
\stopcode
The following interface construct overloads the {\em plus\/} symbol
with the \keyw{vadd} function:
\startcode
\quad \=\quad \kill
\keywb{interface operator}(+) \\
\> \keywb{module procedure} vadd \\
\keywb{end interface}
\stopcode
The same mechanism can be useful for {\em intrinsic\/} functions
and subroutine. For example, the intrinsic function \keyw{abs()}
can be overloaded to calculate the norm of the \keyw{type(vector)}
\startcode
\quad \=\phantom{\keywb{type}(vector), \keywb{intent}(in}, \=\quad \kill
\keywb{function} norm(v) \\
\> \keywb{type}(vector), \keywb{intent}(in) \> :: v \\
\> \keywb{real} \> :: norm \\
\> norm = sqrt(v\%x**2 + v\%y**2) \\
\keywb{end function} norm
\stopcode
Note that the return value is of type \keyw{real}. Other functions
may return the type \keyw{vector}. Again, an interface is needed to
overload \keyw{abs()}
\startcode
\quad \=\quad \kill
\keywb{interface} abs \\
\> \keywb{module procedure} norm \\
\keywb{end interface}
\stopcode
For built-in data types \FORTRANN{} is instructed to
perform array arithmetic, i.e.\
\startcode
\quad \=\quad \kill
\keywb{integer}, dimension(1:10) :: a, b \\
b = abs(a)
\stopcode
is a compact form equivalent to writing:
\startcode
\quad \=\quad \kill
\keywb{integer}, dimension(1:10) :: a, b \\
\keywb{do} i=1, 10 \\
\> b(i) = abs(a(i)) \\
\keywb{enddo}
\stopcode
If we want to do the same with a derived data type or a user defined
function, the function must be given the keyword \keyw{elemental}
introduced in \FORTRANF{}
\startcode
\quad \=\phantom{\keywb{type}(vector), \keywb{intent}(in)} \=\quad \kill
\keywb{elemental function} norm(v) \\
\> \keywb{type}(vector), \keywb{intent}(in) \> :: v \\
\> \keywb{real} \> :: norm \\
\> norm = sqrt(v\%x**2 + v\%y**2) \\
\keywb{end function} norm
\stopcode
This enables the following notation, making array arithmetic available
to the overloaded \keyw{abs()} function:
\startcode
\quad \=\quad \kill
\keywb{type(vector)}, dimension(1:10) :: a, b \\
b = abs(a)
\stopcode

\section{Usage}
\ADF{} constitutes a module that
is written in ISO \FORTRANF{} and should be compatible with any
standard conforming compiler. A new derived type is introduced
in \ADF, namely \keyw{type(ADF95\_dpr)}, which lays out the
memory structure to hold the value and its first derivatives.
All \FORTRANF{} operators and intrinsic functions are implemented
for this type. The
user can choose a \keyw{kind} and must specify
\keyw{LDsize} which is a number less or equal the number of
independent variables. Some additional user functions are provided, to
specify a variable as independent, to
make extraction of values and derivatives easy and to find the
optimal value for \keyw{LDsize}.

\subsection{A first example}
Consider we would like to calculate the
first derivative with respect to the
independent variable \keyw{x}
of the following \FORTRAN{} expression
\startcode
\quad \=\quad \kill
\keywb{real} :: f, x \\
x = 1.0 \\
f = sin(x**2)
\stopcode
The only changes required by the user are to make the module
\keyw{mod\_adf95} available, change the data type \keyw{real}
to \keyw{type(ADF95\_dpr)} and call the routine
\keyw{ADF95\_independent()} to set the independent variables (second argument)
and initial values (third argument):
\startcode
\quad \=\quad \kill
\keywb{use} mod_adf95 \\
\keywb{type}(ADF95\_dpr) :: f, x \\
\keywb{call} ADF95\_independent(1,x,1.0) \\
f = sin(x**2)
\stopcode
Note that the code containing the mathematical evaluation
is not changed. This convenient property is retained also for
arrays, i.e.\
\startcode
\quad \=\quad \kill
\keywb{use} mod_adf95 \\
\keywb{type}(ADF95\_dpr), dimension(1:2) :: f, x \\
\keywb{call} ADF95\_independent((/1,2/),x,(/1.0,5.0/)) \\
f = sin(x**2)
\stopcode
Each independent variable must be given a unique index.
User functions to extract the value and the derivatives from the
last expression are provided and discussed in detail in
Section~\ref{sec:details}.
\begin{figure}[b]
\startcode
\quad \=\quad \=\phantom{\keywb{real}, intent(in)} \=\quad \kill
module my_module \\
\> \keywb{interface} my_func \\
\>\> \keywb{module procedure} my_func1 \\
\> \keywb{end interface} \\
\keywb{contains} \\
\> \keywb{elemental function} my_func1(x, y) \keywb{result}(f) \\
\>\> \keywb{implicit none} \\
\>\> \keywb{real}, intent(in) \> :: x, y \\
\>\> \keywb{real} \> :: f \\
\>\> f = sqrt(abs(x**2-y**2)) + 1.0 \\
\> \keywb{end function} my_func1 \\
\keywb{end module} \\
\\
\keywb{program} original \\
\> \keywb{use} my_module \\
\> \keywb{implicit none} \\
\> \keywb{real}, dimension(1:10) :: fv, gv, xv, yv \\
\> \keywb{integer} :: i \\
\\
\> xv(1:10) = real((/(i,i=1,10)/))**2 \\
\> yv(1:10) = 1.0 / real((/(i,i=1,10)/)) \\
\\   
\> fv(1:10)   = my_func(xv(1:10),yv(1:10))**2 \\
\> gv(1)      = sum(fv(1:10)) \\
\> gv(3: 9:2) = log(fv(4:10:2)**2) \\
\> gv(2:10:2) = exp(1.0/(fv(1: 9:2)**2)) \\
\keywb{end program} original
\stopcode
\caption{Code segment to be changed to allow for automatic
differentiation.\label{fig:prg1}}
\end{figure}

\subsection{A second example}
\begin{figure}[b]
\startcode
\quad \=\quad \=\quad \kill
module my_module \\
\> \keywb{interface} my_func \\
\>\> \keywb{module procedure} my_func1, my_func1_ADF \\
\> \keywb{end interface} \\
\keywb{contains} \\
\> \keywb{elemental function} my_func1(x, y) \keywb{result}(f) \\
\>\> \keywb{implicit none} \\
\>\> \keywb{real}, intent(in) :: x, y \\
\>\> \keywb{real}             :: f \\
\>\> f = sqrt(abs(x**2-y**2)) + 1.0 \\
\>\> \keywb{end function} my_func1 \\
\\
\> \keywb{elemental function} my_func1_ADF(x, y) \keywb{result}(f) \\
\>\> \keywb{use} mod_adf95 \\
\>\> \keywb{implicit none} \\
\>\> \keywb{type}(ADF95_dpr), intent(in) :: x, y \\
\>\> \keywb{type}(ADF95_dpr)             :: f \\
\>\> f = sqrt(abs(x**2-y**2)) + 1.0 \\
\> \keywb{end function} my_func1_ADF \\
\keywb{end module} \\
\\
\keywb{program} original \\
\> \keywb{use} mod_adf95 \\
\> \keywb{use} my_module \\
\> \keywb{implicit none} \\
\> \keywb{type}(ADF95_dpr), dimension(1:10) :: fv, gv, xv, yv \\
\> \keywb{integer} :: i \\
\\
\> call ADF95_independent((/(i,i =1,10)/),xv(1:10),real((/(i,i=1,10)/))**2) \\
\> call ADF95_independent((/(i,i=11,20)/),yv(1:10),1.0/real((/(i,i=1,10)/))**2) \\
\\   
\> fv(1:10)   = my_func(xv(1:10),yv(1:10))**2 \\
\> gv(1)      = sum(fv(1:10)) \\
\> gv(3: 9:2) = log(fv(4:10:2)**2) \\
\> gv(2:10:2) = exp(1.0/(fv(1: 9:2)**2)) \\
\keywb{end program} original
\stopcode
\caption{Modifications of code presented in Fig.~\ref{fig:prg1}
to allow for automatic differentiation. Adding a new
module procedure can save a lot of execution time when
calculation of derivatives are not needed.\label{fig:prg2}}
\end{figure}
A more comprehensive example demonstrates the changes to be
made when function and subroutine calls are involved.
Extensive use of array arithmetic is made to demonstrate this
capability of \ADF. Consider we would like
to calculate the derivatives of the original
code segment shown in Fig.~\ref{fig:prg1}.

As before, the module \keyw{mod\_adf95}
must be made available within all scopes where derivatives should
be calculated. Next, \keyw{ADF95\_independent()} must be called
to specify the independent variables and initial values.
All independent and dependent variables must be changed to
\keyw{type(ADF95\_dpr)}. Since the function \keyw{my\_func1}
may also be called in a context in which the original
version is expected, it is better to add a new
\keyw{module procedure} to it (Fig.~\ref{fig:prg2}).

It is good practice to add a new function to every existing
one that may be needed for differentiation and combine them in a
\keyw{module procedure}. Thus, differentiation is only performed,
when it is actually needed. Purely value oriented operations
will choose the matching \keyw{module procedure} thereby omitting
unnecessary differentiations. Even more importantly,
this approach omits time consuming memory allocations that
would be otherwise necessary because of overloaded function
calls with the data \keyw{type(ADF95\_dpr)}. Thus,
adding \keyw{module procedures} can save a lot of
execution time, even more so if the data structure of
\keyw{type(ADF95\_dpr)} is large due to many independent variables.
The authors of \AUTODERIV{} implemented
a switch which signals when derivatives are to be calculated. However,
this approach is not very efficient compared to
adding new \keyw{module procedures}, mainly because of the
unnecessary memory allocations described above.

\subsection{Full Description}
The modifications needed for an existing \FORTRAN{}
program in order to evaluate first derivatives with \ADF{} are
as follows:\\

In module \keyw{mod\_adf95.f90}:
\begin{enumerate}
\item For a first guess, the the constant parameter \keyw{LDsize} should
be set to the number of independent variables. The best performance is
achieved with the smallest \keyw{LDsize} possible for the 
problem to be differentiated. \keyw{LDsize} is the maximum number
of dependencies from other independent variables.
In many applications, this number
is much smaller than the total number of independent variables themselves.
To illustrate this point further, consider the following example:\\
\startcode
\quad \=\quad \kill
\keywb{call} ADF95_independent((/(i,i=1,10)/),x(1:10),1.0) \\
f(2:9) = x(3:10) - 2 * x(2:9) + x(1:8) \\
\stopcode
where the $x_i$ are $10$ independent variables. Since all $f_i$ are only
functions of three independent variables,
i.e.\ $f_i = f_i(x_{i-1},x_{i},x_{i+1})$, the best choice for
\keyw{LDsize} is $3$. Guessing
the best value for \keyw{LDsize} is almost impossible
for large and complex codes. Therefore,
the user function \keyw{ADF95\_fillin()} is provided to
inquire about the optimal value for \keyw{LDsize}.\\
\item If necessary, the {\em kind\/} parameter \keyw{dpr}
needs to be changed to
the appropriate value for the input variables. The default is
to have \keyw{dpr = KIND(1.0D0)} which is \keyw{double precision}.
If the code uses single precision only, one might like to change
this kind to single precision. Other kind parameters provided
for mixed mode arithmetic, i.e.\ \keyw{spr} and \keyw{ipr}, 
can also be changed. Currently, \ADF{} supports all expressions among
variables of types \keyw{real(dpr)}, \keyw{real(spr)},
and \keyw{integer(ipr)}.
\end{enumerate}
In the user's code; in all scopes where derivatives should be calculated:
\begin{enumerate}
\item Make \keyw{mod\_adf95} accessible through \keyw{use}. Name
clashes with local entities can be avoided by renaming the
few public variables. For example, \keyw{use mod, newname => oldname}
imports the variable \keyw{oldname} from module \keyw{mod}
under the new name \keyw{newname}. The public entities of \ADF{}
are \keyw{ADF95\_independent}, \keyw{ADF95\_value},
\keyw{ADF95\_deriv}, \keyw{ADF95\_fillin} and \keyw{type(ADF95\_dpr)}.
In addition, all operators and many \FFN{} intrinsic functions are
public.\\
\item All independent and dependent, as well as any
intermediate (dependent) variables must be declared
as \keyw{type(ADF95\_dpr)}. If the mathematical expressions
are provided in functions and subroutines, it is advisable to
construct an interface and add a new module procedure to the
existing function or subroutine only with different input and output
variables of \keyw{type(ADF95\_dpr)}. For codes with many
expressions, the {\em include\/}
statement can be used to omit extensive
code repetition. Implicit typing is
permissible, but highly discouraged since it has the
side-effect of declaring constants and other variables as
\keyw{type(ADF95\_dpr)} that are not related to the differentiation
process, thereby wasting memory and execution speed.\\
\item The independent variables must be declared in the
parent scope of the differentiation process. This is easily
done by calling the user function \keyw{ADF95\_independent} which
provides
a method to assign an index and a value
to each independent variable. All indices must be unique,
the lowest index must be one and subsequent indices
should not differ by more than one with respect to
the previous index. However,
the index order is arbitrary.\\
\item After the final assignment to the dependent variable,
the real value of it can be extracted by calling \keyw{ADF95\_value(f)}
where $f$ is a variable of \keyw{type(ADF95\_dpr)}. The
first derivative of $f$ with respect to the independent variable with
index $i$ can be extracted by calling \keyw{ADF95\_deriv(f,i)}.
\end{enumerate}

Following these rules, changes are neither required in the argument list
of any function or subroutine nor
in any statement or mathematical expression. Almost all modifications
can be constrained to interfaces and the declarations
of variables within the interface block and/or the module procedures.

\subsection{Special Cases}
Some potential difficulties may arise from old
\FORTRANS{} programming style and from kind conversions.
\begin{itemize}
\item The use of \keyw{common} blocks and \keyw{equivalence}
statements is still widespread, although their use is
discouraged by the current standard and should be
replaced by automatic arrays, allocatable arrays, pointers and
the \keyw{transfer} statement. Passive variables, such as constants,
pose no problems. However, any active variable, that is passed
through a \keyw{common} block or that is \keyw{equivalence}d should
be renamed and duplicated as follows:\\
\startcode
\quad \=\quad \kill
\keywb{real} :: constant ! no problems \\
\keywb{real} :: x, y, z  ! active variables \\
\\
\keywb{equivalence}(y,z) \\
\keywb{common} /block/ constant, x \\
\\
!-------------\\
! use x, y, z \\
!-------------\\

\stopcode
should become\\
\startcode
\quad \=\quad \kill
\keywb{real} :: constant    ! no problems \\
\keywb{real} :: x_, y_, z_  ! rename variables \\
\keywb{type}(ADF95_dpr) :: x, y, z \\
\\
\keywb{equivalence}(y_,z_) \\
\keywb{common} /block/ constant, x_ \\
\\
\keywb{call} ADF95_independent(1, x, x_) \\
\keywb{call} ADF95_independent(2, y, y_) \\
\\
! y and z not independent\\
z = y\\
\\
!-------------\\
! use x, y, z \\
!-------------\\
\\
! at the end of the routine\\
x_ = ADF95_value(x)\\
y_ = ADF95_value(y)\\
z_ = ADF95_value(z)\\

\stopcode
\item In \FORTRANS{} the use of double precision versions of
trigonometric and other mathematical functions was encouraged, i.e.\
\keyw{dsin(x)} was used for \keyw{double precision} types.
In \ADF{} only the standard conforming generic names are implemented.
The user must
therefore change all occurrences, for example, of \keyw{dsin(x)} to
\keyw{sin(x)}.\\
\item Type conversions from one kind to another is
not permissible for variables of \keyw{type(ADF95\_dpr)}, since only
one type is implemented. Actually, it is not possible
to construct a user defined function in
\FORTRANF{} that can return values with different kinds.
Thus, expressions such as \keyw{y=real(x,1.d0)}
or the obsolete \FORTRANS{} expression \keyw{y=dble(x)} must be
omitted.
\end{itemize}

\subsection{Output Verification}
\begin{figure}
\startcode
\quad \=\phantom{\keywb{type}(ADF95_dpr)} \=\quad \kill
\keywb{program} verify_out \\
\> \keywb{use} mod_adf95 \\
\> \keywb{implicit none} \\
\> \keywb{type}(ADF95_dpr)\>, dimension(1:2) :: f , x \\
\> \keywb{real}(dpr)      \>, dimension(1:2) :: fp, xp\\
\\
\> xp = (/1.0,5.0/) \\
\> call ADF95_independent((/1,2/),x,xp) \\
\\
\> f  = sin(x**2)\\
\> fp = 2.0_dpr*xp*cos(xp**2)\\
\\
\> write(*,'(A,2(ES25.15))') "x array =", ADF95_value(x)\\
\> write(*,'(A,2(ES25.15))') "f array =", ADF95_value(f)\\
\> write(*,*) "***ADF95:"\\
\> write(*,'(A,2(ES25.15))') "df/dx1  =", ADF95_deriv(f,1)\\
\> write(*,'(A,2(ES25.15))') "df/dx2  =", ADF95_deriv(f,2)\\
\> write(*,*) "***Analytic:"\\
\> write(*,'(A,2(ES25.15))') "df/dx1  =", fp(1)  , 0.0_dpr\\
\> write(*,'(A,2(ES25.15))') "df/dx2  =", 0.0_dpr, fp(2)\\
\keywb{end program} verify_out
\stopcode
\caption{Code segment to verify correct compilation of \ADF{}.
\label{fig:ver1}}
\end{figure}
To be able to test for successful compilation of \ADF{} and verify
the correct solution I provide a simple example
in Fig.~\ref{fig:ver1}. Running the executable should yield
the following output:
\startcode
x array =    1.000000000000000E+00    5.000000000000000E+00\\
f array =    8.414709848078965E-01   -1.323517500977730E-01\\
 ***ADF95:\\
df/dx1  =    1.080604611736280E+00    0.000000000000000E+00\\
df/dx2  =    0.000000000000000E+00    9.912028118634735E+00\\
 ***Analytic:\\
df/dx1  =    1.080604611736280E+00    0.000000000000000E+00\\
df/dx2  =    0.000000000000000E+00    9.912028118634735E+00
\stopcode
The last digits may vary depending on the system architecture, but
outputs from \ADF{} when compared to the analytic approach (last
two lines of output) must be identical.

\section{Implementation}
\ADF{} is a \FORTRANF{} module containing functions that overload
all operators and all appropriate \FORTRANNF{} intrinsic functions
for the new derived data \keyw{type(ADF95\_dpr)}. The data structure
of \keyw{type(ADF95\_dpr)} is simple: it holds one entry for
the value, \keyw{LDsize} entries for the values of derivatives
and \keyw{LDsize+1} values for indices:
\startcode
\quad \=\quad \kill
\keywb{type} ADF95_dpr \\
\> \keywb{real}   (dpr)                      :: value  = 0.0_dpr \\
\> \keywb{real}   (dpr), dimension(1:LDsize) :: deriv  = 0.0_dpr \\ 
\> \keywb{integer}(ipr), dimension(0:LDsize) :: index  = 0_ipr   \\
\keywb{end type} ADF95_dpr
\stopcode
The entry for \keyw{index(0)} is reserved for the current number 
of non-zero derivatives. The values for the indices correspond
to the index of the independent variable with respect
to which the derivative is taken.
For illustration, consider that the variable $f$ is a
function of the independent variable $x$ and further that
$f(x) = 1, f'(x) = 2$. The representation on \keyw{type(ADF95\_dpr)}
would be:
\startcode
\quad \=\quad \kill \\
f\%value    = 1.0  ! f(x) = 1 \\
f\%index(0) = 1    ! number of derivatives \\
f\%index(1) = 1    ! unique index of x \\
f\%deriv(1) = 2.0  ! f'(x) = 2
\stopcode
This indexing technique leads to compact storage of derivatives and
--- since \keyw{LDsize} is in many applications much smaller than
the total number of independent variables  --- to an economical
memory use which is rewarded by faster execution speeds.

Note that \keyw{LDsize} must be chosen before the
compilation of the program and that all variables of
\keyw{type(ADF95\_dpr)} allocate the same
amount of memory. Since not all of those variables actually need
\keyw{LDsize} entries, memory and execution speed is wasted.
Dynamic memory allocation could be used through the \keyw{allocatable}
keyword which is nowadays supported also for derived types
by many \FORTRANF{} compilers and that is part of the new \FORTRAND{}
standard \cite{fortran03}. However, all my actual implementations
resulted in considerably slower execution speeds in practical
applications. This is probably due to the overhead needed to
decide when new memory must be allocated/deallocated and, more likely,
because of the time needed for the allocation process and for
the access to the resulting scattered memory locations. These
findings may well change with future
compilers\footnote{Tests were only performed with the Lahey/Fujitsu F95 compiler.}
and further research in this direction is needed.

Due to the overloading of operators and intrinsic functions the
compiler generates code for the evaluation of the value and
the numerical derivatives according to the chain rule of
differentiation whenever operations between \keyw{type(ADF95\_dpr)}
are encountered. Mixed mode arithmetic is also supported through
additional module procedures provided in \ADF{}.
For example, with variable \keyw{a} of type \keyw{real(dpr)} and
variables \keyw{b}, \keyw{c} of \keyw{type(ADF95\_dpr)}
the compiler parsing the statement
\begin{equation}
  c = a \cdot b 
\end{equation}
generates code for the value and its derivatives:
\begin{eqnarray}
  c &=& a \cdot b \\[3mm]
\frac{\partial c}{\partial q_i} &=& a\, \frac{\partial b}{\partial q_i}
\end{eqnarray}
This is accomplished technically by the following function
that overloads \keyw{operator(*)}:
\startcode
\quad \=\phantom{\keywb{type}(ADF95_dpr)} \=\phantom{,intent(in)} \=\quad \kill
\keywb{elemental} function multiply(a, b) result(f) \\
\> \keywb{use} mod_precision \\
\> \keywb{implicit none} \\
\> \keywb{real}(dpr)       \>, intent(in) \> :: a \\
\> \keywb{type}(ADF95_dpr) \>, intent(in) \> :: b \\
\> \keywb{type}(ADF95_dpr) \>\> :: f \\
\> \keywb{integer}(ipr)    \>\> :: lenb \\
\\
\> lenb    = b\%index(0) \\
\> f\%value = a * b\%value \\
\\
\> f\%deriv(1:lenb) = a * b\%deriv(1:lenb) \\
\> f\%index(0:lenb) =     b\%index(0:lenb) \\
\keywb{end function} multiply
\stopcode
The only parameters defined in the module are the kinds
of the components in \keyw{type(ADF95\_dpr)}, i.e.\ \keyw{dpr},
and those needed for mixed mode arithmetic, i.e.\ \keyw{spr} and
\keyw{ipr}. These parameters can be changed to extend the
precisions. In order to avoid clashes in overloading resolution,
\keyw{dpr} and \keyw{spr} must have different values.
Currently, \FORTRANF{}
does not provide a mechanism to utilise implicit promotions
from one derived type to another nor does it allow to define
conversions between derived types. Therefore, all procedures had
to be written into supported types. Also note, that complex
variables are not supported.

\subsection{User functions}
\label{sec:details}
The mathematically
independent variables must be specified at run-time, therefore
\ADF{} provides the user function \keyw{ADF95\_independent}.
The routine accepts three arguments, the variable, a value
and an integer index that must be unique. 
This routine assigns the value and sets the derivative
with respect to itself to \keyw{1.0\_dpr}. Its interface is:
\startcode
\quad \=\phantom{\keywb{type}(ADF95_dpr)} \=\phantom{,intent(inout)} \=\quad \kill
\keywb{elemental subroutine} ADF95_independent(i,x,val) \\
\> \keywb{integer}(ipr)    \>, intent(in)   \> :: i\\
\> \keywb{type}(ADF95_dpr) \>, intent(inout)\> :: x\\
\> \keywb{real}(dpr)       \>, intent(in)   \> :: val\\
\keywb{end subroutine} ADF95_independent
\stopcode
Three different versions are overloaded such that
\keyw{ADF95\_independent} accepts values of the types
\keyw{real(dpr)}, \keyw{real(spr)} and \keyw{integer(ipr)}.
The value of the variable of \keyw{type(ADF95\_dpr)} can be
extracted by calling \keyw{ADF95\_value}. Its function
interface is:
\startcode
\quad \=\phantom{\keywb{type}(ADF95_dpr), intent(in)} \=\quad \kill
\keywb{elemental function} ADF95_value(x) result(f) \\
\> \keywb{type}(ADF95_dpr), intent(in) \> :: x \\
\> \keywb{real}(dpr) \> :: f \\
\keywb{end function} ADF95_value
\stopcode
Similarly, the function \keyw{ADF95\_deriv}
is provided to extract the derivatives. In
addition to the \keyw{type(ADF95\_dpr)} a second argument
is expected, the index of the independent variable to which
respect the derivative is taken:
\startcode
\quad \=\phantom{\keywb{type}(ADF95_dpr)} \=\phantom{,intent(in)} \=\quad \kill
\keywb{elemental function} ADF95_deriv(x, i) result(df) \\
\> \keywb{type}(ADF95_dpr) \>, intent(in) \> :: x \\
\> \keywb{integer}(ipr) \>, intent(in) \> :: i \\
\> \keywb{real}(dpr) \>\> :: df \\
\keywb{end function} ADF95_deriv
\stopcode
Two additional user routines are provided for convenience. The
first subroutine, \keyw{ADF95\_jacobian}, expects an array of
\keyw{type(ADF95\_dpr)} and returns three arrays
containing derivatives and indices. For example, \keyw{df(k)}
is the derivative of
$\partial \keyw{f(ir(k))} / \partial \keyw{x(ic(k))}$.
The integer return value \keyw{nz}, contains the number
of non-zero entries in \keyw{df} or a negative value if the
array size of \keyw{df}, \keyw{ic} or \keyw{ir} is
not sufficiently large:
\startcode
\quad \=\phantom{\keywb{type}(ADF95_dpr)} \=\phantom{,dimension(:)} \=\phantom{, intent(in)} \=\quad \kill
\keywb{pure subroutine} ADF95_jacobian(f, df, ir, ic, nz) \\
\> \keywb{type}(ADF95_dpr) \>, dimension(:) \>, intent(in)  \> :: f \\
\> \keywb{real}(dpr) \>, dimension(:) \>, intent(out) \> :: df \\
\> \keywb{integer}(ipr) \>, dimension(:) \>, intent(out) \> :: ir, ic \\
\> \keywb{integer}(ipr) \>\>, intent(out) \> :: nz \\
\keywb{end subroutine} ADF95_jacobian
\stopcode
Finally, the function \keyw{ADF95\_fillin} inquires about
the optimal value for \keyw{LDsize}. Its input argument
is the the (array of) variables of the final assignment statement.
Two optional integer arguments \keyw{ml} and \keyw{mu}
are returned with the number of non-zero sub-diagonals
and/or super-diagonals, respectively.
\startcode
\quad \=\phantom{\keywb{type}(ADF95_dpr)} \=\phantom{, dimension(:)} \=\phantom{, intent(out)} \=\quad \kill
\keywb{pure subroutine} ADF95_fillin(f, LDsize_opt, ml, mu) \\
\> \keywb{type}(ADF95_dpr) \>, dimension(:) \>, intent(in) \> :: f \\
\> \keywb{integer}(ipr) \>\>, intent(out) \> :: LDsize_opt \\
\> \keywb{integer}(ipr) \>, optional \>, intent(out) \> :: ml, mu \\
\keywb{end subroutine} ADF95_fillin
\stopcode
It must be stressed that \keyw{ADF95\_fillin} returns only
the correct number for \keyw{LDsize} if \ADF{} was
compiled with a sufficiently large 
\keyw{LDsize} in the first place.
If a sensible initial guess for \keyw{LDsize} is not possible,
\keyw{LDsize} should be set to the total number of independent
variables before compiling \ADF{}.
Next inquire about the best value for
\keyw{LDsize} by calling \keyw{ADF95\_fillin} and set
it to the inquired value. Finally re-compile \ADF{}.

\subsection{Supported \FORTRANNF{} intrinsics}
Great care has been taken to overload all \FORTRANNF{} intrinsic
functions and built-in operators
for the new data \keyw{type(ADF95\_dpr)} whenever meaningful.
Fully supported are the following routines
including the capability to accept and return conformable arrays:
\keyw{abs},
\keyw{atan},
\keyw{cos},
\keyw{cosh},
\keyw{digits},
\keyw{dim},
\keyw{dot\_product},
\keyw{epsilon},
\keyw{exp},
\keyw{exponent},
\keyw{fraction},
\keyw{huge},
\keyw{kind},
\keyw{log},
\keyw{log10},
\keyw{matmul},
\keyw{maxexponent},
\keyw{minexponent},
\keyw{mod}, 
\keyw{modulo}, 
\keyw{nearest},
\keyw{precision},
\keyw{radix},
\keyw{range},
\keyw{rrspacing},
\keyw{scale},
\keyw{set\_exponent},
\keyw{sign}, 
\keyw{sin},
\keyw{sinh},
\keyw{spacing},
\keyw{tan},
\keyw{tanh},
\keyw{tiny}.
For some others, exactly the same behaviour as for built-in functions
cannot be overloaded. These limitations are described in the next section.

\subsection{Implementation details of \keyw{tanh}}
The derivative of \keyw{tanh(x)} with respect to \keyw{x}
is given by \keyw{1.0/cosh(x)**2}. For increasing \keyw{x}
the hyperbolic cosine grows beyond all limits. Thus, \keyw{cosh}
produces a floating point exception for large \keyw{x}. To
circumvent this situation, the derivative could be
calculated from the mathematically equivalent form
\keyw{1.0-tanh(x)**2} as done in \cite{stamatiadis00}. This
formula avoids floating point exceptions but due to finite
computer precision the result is resolved to zero for relatively
small \keyw{x} rather than to a finite number.

A better implementation is chosen in \ADF{}. The formula
\keyw{1.0/cosh(x)**2} is used for \keyw{abs(x) < 2.0*range(x)}
in which case cosh can be calculated. For larger \keyw{abs(x)} the
derivative is approximated with \keyw{4.0*exp(-2.0*abs(x))}. Thus
a finite number is returned which can be as low as the smallest
number that is representable in the current data model without
being hampered by finite precision. For even larger \keyw{x}
the derivative is resolved to zero.

\subsection{Limitations}
The built-in intrinsic functions 
\keyw{aint},
\keyw{anint},
\keyw{ceiling},
\keyw{floor},
\keyw{int} and
\keyw{nint}
can be called with an optional {\em kind\/} parameter such that
the return value has the same kind. Since \FORTRANF{} does
not allow the kind of a derived type's component to be chosen
when the derived type is used, this functionality cannot be implemented.
However, this situation will change in the near future with
the advent of \FORTRAND{} \cite{fortran03}.

A similar problem arises with functions that accept arrays of different
rank. Since the rank cannot be chosen dynamically in user defined
functions, a new module procedure must be added for all possible
ranks. \ADF{} accepts only arrays of rank one in these instances.
The functions
\keyw{maxloc},
\keyw{maxval},
\keyw{minloc},
\keyw{minval},
\keyw{product} and
\keyw{sum}
are affected by this restriction. For the same reason, the optional
parameter \keyw{dim} is not supported for these functions.

The functions \keyw{max} and \keyw{min} accept a variable
number of arguments for built-in types. This cannot be implemented
either. A simple work-around to this deficiency is 
to change all instances in which more than two arguments are used
from \keyw{max(v1, v2,\ldots, vn)} to
\keyw{max(\ldots max(v1,v2),\ldots, vn)}.

\subsection{Undefined derivatives}
\label{sec:notdefined}
Any mathematical operation between
values of \keyw{type(ADF95\_dpr)},
that is forbidden (e.g., division by zero)
is treated exactly the same as for built-in types and
produces floating point exceptions. No additional
coding is needed in these instances. However, in some
functions a situation can occur where the operation
on the value is permissible while the derivative is not defined.

Serious problems of this kind
arise in cases where the function is
not mathematically differentiable. For example, the derivative
of \keyw{abs(f(x))} at \keyw{f(x)=0}, \keyw{f'(x)$\ne$0}
is not defined,
likewise the derivative of \keyw{sqrt(f(x))} at \keyw{f(x)=0}. 
Undefined situations as such can occur in the functions
\keyw{acos},
\keyw{asin},
\keyw{atan2},
\keyw{max},
\keyw{maxval},
\keyw{min},
\keyw{minval} and
\keyw{sqrt}.
Divisions by zero return \keyw{Inf} (Infinity) or, depending on the
compiler options, a floating point exception. In all other
cases \ADF{} is instructed to return \keyw{-sqrt(-1.0)} which
yields, depending on the system and compiler options,
either \keyw{NaN} (Not a Number) or a again a
floating point exception. 
Note that computing the analytic derivatives by other means
would lead to the same undefined situations.

In \AUTODERIV{}, these occurrences are arbitrarily resolved to zero
which is mathematically incorrect.
The approach of \ADF{} has the advantage that the user is being
notified that an illegal mathematical operation has been performed,
pointing him to the location where his code needs rethinking.

\section{Tests}

\subsection{Verifying the solution}
In order to test the correctness of the solution
calculated with \ADF{}, numerous comparisons between
\ADF{} and \AUTODERIV{} including {\em all\/} overloaded operators
and functions (as well as combinations among them) were performed.
These comparisons revealed an error in the function \keyw{fraction}
of \AUTODERIV{}: the return value must be of type \keyw{real}
and not of type \keyw{integer}.

The results of all other operations and functions turned out
to be {\em identical}. Since both modules were developed
completely independently this result is a strong indication for the
correctness of both packages. Nevertheless, it is almost impossible
to cover and compare all possible code branches of both routines,
therefore all tests are inherently incomplete.

As expected, different results were encountered with
\keyw{tanh} and in situations in which undefined
derivatives occur. Those are set to zero in \AUTODERIV{} whereas
\ADF{} sets them to \keyw{NaN} (see Section~\ref{sec:notdefined}).

\subsection{Performance and Compiler Comparison}
A number of tests have been performed in order to measure the efficiency
of \ADF{} in comparison to \AUTODERIV{} and in comparison to
analytically computed and hard-coded derivatives. Five up-to-date
\FORTRAN{} compilers for {\em Linux} must demonstrate their efficiency:
{\em Absoft}, {\em Intel}, {\em Lahey/Fujitsu}, {\em NAG\/} and
{\em PGI}. The compiler options have been
chosen to give maximum execution performance (Table~\ref{tab:opts}).
All tests were performed on a
Mobile Intel(R) Pentium(R) 4 processor
at 2.5~GHz, 1~GB of memory,
running on a {\em Linux/RedHat~8} operating system.
\begin{table} 
  \caption{\FORTRANF{} compilers and compiler options used for test runs.}
  \label{tab:opts}
\begin{minipage}{\textwidth}
  \begin{tabular*}{\hsize}{ll}
\hline
Compiler & Options \\
\hline
Absoft Pro FORTRAN Version 8.2a & --O3 --cpu:p7\\[3mm]
Intel(R) Fortran Compiler for 32-bit applications, & --O3 --ipo --static\footnote{Omitted in third example: generated code causes segmentation fault.}\\[-1mm]
Version 8.0 (Package ID: l\_fc\_pc\_8.0.046)       &                    \\[3mm]
Lahey/Fujitsu Fortran 95 Compiler Release L6.20b & --O --tp4 --trap --staticlink \\[3mm]
NAGWare Fortran 95 compiler Release 5.0(322)     & --O4 --Bstatic --unsharedf95  \\[3mm]
Portland Group, Inc. pgf90 5.1-6\footnote{Note: pgf90 is not a \FORTRANF{} compiler.} & --fast --tp piv\\
\hline
  \end{tabular*}
\end{minipage}
\end{table}

As a first example in \cite{stamatiadis00}, the performance of
\AUTODERIV{} is benchmarked by calculating the derivatives
of the Potential Energy Surface (PES) for the HCP molecule,
described in \cite{beck97}. The
code for the calculation of the PES is available from \cite{farantos96}.
This is a realistic example, but only with three independent
variables and its main purpose is to test \ADF{} with the
exact same piece of code on which \AUTODERIV{} was tested.
The PES code is simple enough that the calculation of
first derivatives ``by hand" is still feasible and I have done
this in order to allow comparison with the automatic
differentiation approach. Table~\ref{tab:hcp} summarises
the results of the HCP example for different methods
and compilers. The variable \keyw{LDsize} had to be set
to $3$ in \ADF{}. The time was measured with the \FORTRAN{}
routine \keyw{system\_clock}.
\begin{table}[b] 
  \caption{Derivatives of the Potential Energy Surface of a HCP molecule.
Time averaged over $10^6$ evaluations and quoted in $\mu$s.}
  \label{tab:hcp}
\begin{minipage}{\textwidth}
  \begin{tabular*}{\hsize}{lrlrrrr}
\hline
Compiler & \multicolumn{3}{l}{Execution time [$\mu$s]} &\multicolumn{3}{l}{Memory usage [kBytes]} \\
         & analytic & \ADF{} & \AUTODERIVS{} & analytic & \ADF{} & \AUTODERIVS{} \\
\hline
Absoft        & $3.6$ & $\phantom{ADF}32$ & $108$ & \multirow{5}{1.5cm}{\hspace*{-1cm}$\left. \phantom{\frac{\frac{\frac{\int A}{\frac{\int a}{\frac{A}{A}}}}{\int A}}{\int A}} \right\}\quad 2.0$}  & \multirow{5}{1.0cm}{$3.3$} & \multirow{5}{1.2cm}{$5.5$} \\
Intel         & $3.7$ & $\phantom{ADF}11$ &  $42$ & & & \\
Lahey/Fujitsu & $1.5$ & $\phantom{ADF}19$ &  $93$ & & & \\
NAG           & $5.3$ & $\phantom{ADF}30$ & $159$ & & & \\
PGI           & $7.3$ & $\phantom{ADF}19$\footnote{\ADF{} was stripped of its \FORTRANF{} features in order to make it run with the PGI compiler.} & $714$ & & & \\
\hline
  \end{tabular*}
\end{minipage}
\end{table}

For three out of five compilers \ADF{} is only a factor of $3$--$6$ slower
compared to the direct analytic computation. Furthermore,
\ADF{} is about a factor of four
faster than \AUTODERIV{} regardless of the compiler chosen. This
is quite surprising, because the advantages of the indexing
method do not show up in systems where \keyw{LDsize} is small or
where it is equal to the number of independent variables.
One reason might stem from the additional memory that must be allocated
in \AUTODERIV{} to hold the second derivatives.
Extensive use of function calls in  \AUTODERIV{} may also produce
additional overhead, unless the compiler is capable of inlining
code properly.

My second example is one from astrophysics. A nuclear
fusion network with $14$ nuclei is operating within every 10 different
temperature/density shells. This corresponds to $140$ independent
variables altogether, but since
only the network is coupled, \keyw{LDsize} can be set to $14$. Thus
this example exploits the advantages of \ADF{} as can be seen in
Table~\ref{tab:ntw}.
\begin{table} 
  \caption{Derivatives for nuclear reaction network example. The
simulation consists of 10 shells with 14 nuclei each which amounts
to 140 independent variables.}
  \label{tab:ntw}
\begin{minipage}{\textwidth}
  \begin{tabular*}{\hsize}{lrlrrrr}
\hline
Compiler & \multicolumn{3}{l}{Execution time [s]} &\multicolumn{3}{l}{Memory usage [kBytes]} \\
         & analytic & \ADF{} & \AUTODERIVS{} & analytic & \ADF{} & \AUTODERIVS{}                 \\
\hline
Absoft        & $1.6$ & \phantom{ADF}$7.3$\footnote{The Absoft compiler does not behave conforming to ISO \FORTRANF{} . ADF95{} had to be altered to make it run with this compiler.} & $5090$ & \multirow{4}{1.5cm}{\hspace*{-1cm}$\left. \phantom{\frac{\frac{\frac{\int A}{\frac{a}{\frac{}{}}}}{\int A}}{\int A}} \right\}\quad 48$}  & \multirow{4}{1.0cm}{$172$} & \multirow{4}{1.2cm}{$74800$} \\
Intel         & $1.0$ & \phantom{ADF}$5.7$ &  $103$ & & & \\
Lahey/Fujitsu & $1.2$ & \phantom{ADF}$7.0$ & $2530$ & & & \\
NAG           & $1.4$ & \phantom{ADF}$7.1$ & $1139$ & & & \\
PGI\footnote{Excluded from test since PGI does not support \FORTRANF{}.} & ---   & \phantom{ADF}---   & ---    & & & \\
\hline
  \end{tabular*}
\end{minipage}
\end{table}

Depending on the compiler \AUTODERIV{} is a factor of $20$--$700$ slower
and uses $400$ times more memory than \ADF{}. Also note that the
performance of \AUTODERIV{} is extremely compiler dependent whereas
\ADF{} is about a factor of $6$ slower compared to the
analytic computation of derivatives regardless of the compiler.

Finally I use the exact same application again, but now
with $1000$ temperature/density shells which amounts
to $14000$ independent variables. This problem cannot
be handled with \AUTODERIV{} any more (Table~\ref{tab:ntw2}). It
can be seen, that as long as \keyw{LDsize} is not changed, the
execution time scales simply with the number of derivatives to
be calculated. Again, computation of
hard-wired analytic derivatives are by a factor of $6$ faster.
The memory requirement of \ADF{} can be easily calculated: A
variable of \keyw{type(ADF95\_dpr)} holds
$(1 + \keyw{LDsize}) * (\keyw{real(dpr)} + \keyw{integer(ipr)})$
numbers. Taking default parameters for \keyw{dpr} and \keyw{ipr}
this amounts to $12\cdot (1 + \keyw{LDsize})$ bytes multiplied
by the number of \keyw{type(ADF95\_dpr)} variables in the program.
\begin{table} 
  \caption{Derivatives for nuclear reaction network example
consisting of 1000 shells with 14 nuclei each which amounts
to 14000 independent variables.}
  \label{tab:ntw2}
\begin{minipage}{\textwidth}
  \begin{tabular*}{\hsize}{lrlrrrr}
\hline
Compiler & \multicolumn{3}{l}{Execution time [s]} &\multicolumn{3}{l}{Memory usage [kBytes]} \\
         & analytic & \ADF{} & \phantom{\AUTODERIVS{}} & analytic & \ADF{} & \phantom{\AUTODERIVS{}}                 \\
\hline
Absoft        & $159$ & \phantom{AD}$715$\footnote{The Absoft compiler does not behave conforming to ISO \FORTRANF{}. ADF95{} had to be altered to make it run with this compiler.} & \phantom{---} & \multirow{4}{1.5cm}{\hspace*{-1cm}$\left. \phantom{\frac{\frac{\frac{\int A}{\frac{a}{\frac{}{}}}}{\int A}}{\int A}} \right\}\quad 3838$}  & \multirow{4}{1.0cm}{$7643$} & \multirow{4}{1.2cm}{\phantom{---}} \\
Intel         & $107$ & \phantom{AD}$566$ & \phantom{---} & & & \\
Lahey/Fujitsu & $109$ & \phantom{AD}$575$ & \phantom{---} & & & \\
NAG           & $131$ & \phantom{AD}$706$ & \phantom{---} & & & \\
PGI\footnote{Excluded from test since PGI does not support \FORTRANF{}.}           & ---   & \phantom{AD}---   & \phantom{---} & & & \\
\hline
  \end{tabular*}
\end{minipage}
\end{table}

\section{Discussion}
If there is need for numerical first derivatives, accurate to
machine precision, which is the case, e.g., for implicit solvers employed
for simulations in all computational sciences, the use of \ADF{}
should be seriously considered. As demonstrated
for realistic examples in this paper, this method
is still about a factor of $6$ slower compared to the method
of hard-wiring the analytically derived first derivatives. Thus, if
maximum performance is demanded, \ADF{} should be employed only
if the part for calculating first derivatives is not
limiting the performance of the entire program. The latter
situation in which the differentiation part is not
crucial to the overall program performance does indeed
occur in state-of-the-art
implicit solvers \cite{ehrig} and little compromise has to be made
when employing \ADF{}.

Apart from these performance considerations, \ADF{} can reduce code
development considerably. In the case of large systems the
analytic differentiation combined with the need of extra coding
is an error-prone task which easily introduces difficult to
find bugs into the program thereby slowing down the development process.
Furthermore, successfully implemented systems are difficult to change,
since it usually requires to alter many equations for calculating
derivatives. Even if one insists on this approach in view of
its performance benefits, \ADF{} can be a convenient tool to
find bugs or verify the solution for calculating derivatives
more quickly. It can be also used to inquire about the structure of
the Jacobian matrix and also to search for non-differentiable
situations within the coded systems of equations which can lead
to the detection of spurious convergence problems.

The disproportionality in performance between
the hard-wiring approach and \ADF{} may well be reduced with
better compiler technology in the future. Although \FORTRANF{}
has been standardised seven years ago, many compilers
are still lacking reliable support for it. {\em PGI\/} does not provide
\FORTRANF{} and the one from {\em Absoft\/} is extremely buggy on the
new features. To a substantially lesser degree, this finding is
also true for the {\em Intel\/} compiler. Throughout the whole study
the {\em Intel\/} compiler produces the fastest executables but
best support for ISO \FORTRANF{} is provided by the compilers
from {\em NAG} and {\em Lahey/Fujitsu}. The latter offers the
best compromise between stable language support and execution speed.
It can be suspected that
there is still room for compiler optimisations when \FORTRANF{}
constructs are involved.

On the other hand, the design of \ADF{} may also be improved upon.
The approach of allocating memory statically leads to some
waste of memory. Dynamic memory allocation might improve
on the performance of \ADF{} but my first tests on this showed, unfortunately,
the opposite effect. Also, the default initialisation of all
entries within \keyw{type(ADF95\_dpr)} is not needed but
the code for the overloaded functions would have been more complicated
otherwise.

With \ADF{} an easy to use automatic differentiation tool is now
available efficient enough worth being employed in
many realistic applications.
\ack
This research has been supported in part by the
{\em Deutsche Forschungsgemeinschaft},
DFG (SFB 439 Galaxies in the Young Universe) and the stipend of
the {\em Elitef\"orderprogramm f\"ur Postdoktoranden der Landesstiftung
Baden-W\"urttemberg}.

\end{document}